\documentclass[twocolumn,showpacs,preprintnumbers,amsmath,amssymb]{revtex4}
\usepackage{amsmath,amssymb,graphics,epsfig,subfigure}
\usepackage{color}

\begin{document}
\newcommand {\nn} {\nonumber}
\renewcommand{\baselinestretch}{1.3}

\title{Black hole entropy emission property}

\author{Shao-Wen Wei \footnote{weishw@lzu.edu.cn},
        Yu-Xiao Liu \footnote{liuyx@lzu.edu.cn}}

\affiliation{Institute of Theoretical Physics, Lanzhou University, Lanzhou 730000, People's Republic of China}

\begin{abstract}
In this work, we examine the entropy emission property of black holes. When the greybody factor is considered, it is found that Schwarzschild black hole is a one-dimensional entropy emitter, which is independent of the spacetime dimension and the spin of the emitted quanta. However, when generalized to other black holes with two or more parameters, the result shows that the one-dimensional entropy emission property will be violated. Thus our result implies that not all black holes behave as one-dimensional entropy emitters.
\end{abstract}



\maketitle

\section{Introduction}
\label{secIntroduction}

Combining quantum mechanics and general relativity, black holes are found to radiate particles characterized by the well-known Hawking temperature \cite{Hawking,Hawkingb}. There are several approaches proposed for obtaining the temperature, for examples the collapse geometry, tunneling process \cite{Parikh}, and the gauge and gravitational anomalies \cite{Robinson,Iso}.

Exploration of the property corresponding to Hawking radiation is an important subject. In 2001, Bekenstein and Mayo \cite{Bekenstein} studied the black hole entropy flow related with the radiation particles at the Hawking temperature, and found that the entropy emission rate for a (3+1)-dimensional Schwarzschild black hole is
 \begin{eqnarray}
 \dot{S}=\left(\frac{\nu^{2}\bar{\Gamma}\pi P}{480 \hbar}\right)^{\frac{1}{2}},
\end{eqnarray}
where the values of $\nu$ and $\bar{\Gamma}$ can be found in Page's work \cite{Page,Pageb}. This result strongly recommends that black hole radiation is completely different from that of a hot body in three-dimensional space, but like a hot body in one-dimensional space. Thus they concluded that Schwarzschild black holes are one-dimensional entropy emitters, and the information flow accompanied by Hawking radiation from a (3+1)-dimensional Schwarzschild black hole is one-dimensional. Therefore, such property can be treated as a complementary statement to the famous ``holographic principle" \cite{Hooft,Susskind}.

Very recently, the one-dimensional nature of the black hole entropy emission rate has been explored with a direct calculation of the entropy emission rate. The first interesting example is the $d$-dimensional Schwarzschild black hole. In Refs.~\cite{Mirza,Hod0}, it was shown that its entropy emission rate has a similar form as that of a one-dimensional hot body, which implies that a Schwarzschild black hole in arbitrary spacetime dimensions is a one-dimensional entropy emitter. This greatly confirms Bekenstein and Mayo's result, and supports the conjecture that \emph{all black holes are one-dimensional entropy emitters}. However, when generalized other types of black holes, the result behaves very differently. For a (2+1)-dimensional non-rotating Banados-Teitelboim-Zanelli (BTZ) black hole, the result \cite{Mirza} shows that the entropy flow or information out of the black hole is three-dimensional. For Lovelock black holes, the result indicates that the channel of the entropy flow is equal to $d$ for odd $d$-dimensional spacetime, and $1+\varepsilon(\Lambda)$ for even $d$-dimensional spacetime \cite{Mirza}. On the other hand, Hod \cite{Hod} noted that the entropy emission rate of a Reissner-Nordstr\"{o}m (RN) black hole characterized by the neutral sector of the Hawking radiation spectra can be studied analytically in the near-extremal regime. Unfortunately, the result indicates that such black hole is not one-dimensional entropy emitter.

Nevertheless, we would like to reexamine the black hole entropy emission property. In the previous works, the study showed that the Schwarzschild black hole is a one-dimensional entropy emitter of boson quanta. However, how about the fermion radiation? The answer is worth searching. On the other hand, when some other parameters are included in, do the black holes still behave as one-dimensional entropy emitters? Motivated by these questions, we would like to explore the entropy emission properties of black holes, even the gravitational effect is considered.

\section{Bekenstein and Mayo's treatment}

Here, we would like to give a brief review of the Bekenstein and Mayo's treatment given in Ref. \cite{Bekenstein}. In flat spacetime, the energy and entropy transmission out of a closed black object with temperature $T$ and area $A$ into  3-dimensional space are
\begin{eqnarray}
 P&=&\frac{\pi^{2}AT^{4}}{120 h^{3}},\\
 \dot{S}&=&\frac{4P}{3T}.
\end{eqnarray}
Combining the above two equations yield
\begin{eqnarray}
 \dot{S}=\frac{2}{3}\bigg(\frac{2\pi^{2}AP^{3}}{15\hbar^{3}}\bigg)^{1/4}.\label{Srate}
\end{eqnarray}
On the other side, if we replace the black object with a black hole, the entropy emitted rate is also in the form (\ref{Srate}). However, the area $A$ and temperature $T$ will not be independent of each other. For a Schwarzschild black hole, we always have
\begin{eqnarray}
 A=16\pi M^{2},\quad T=\frac{\hbar}{8\pi M},
\end{eqnarray}
where $M$ is the black hole mass. Thus, one will obtain \cite{Bekenstein}
\begin{eqnarray}
 \dot{S}=\left(\frac{\nu^{2}\bar{\Gamma}\pi}{480\hbar}\right)^{1/2}\times P^{1/2}.
\end{eqnarray}
As claimed by Bekenstein and Mayo, this result implies that a $d$=4-dimensional Schwarzschild black hole is a one-dimensional entropy emitter.

This result was soon generalized to the higher dimensional Schwarzschild black hole cases (see Refs. \cite{Mirza,Hod0}). The starting point is the generalized Stefan-Boltzmann law, which in $d$-dimensional spacetime is given by \cite{Castro}
\begin{eqnarray}
 P=\sigma_{d}\mathcal{A}T^{d}, \label{pat}
\end{eqnarray}
where $\sigma_{d}$ is the generalized Stefan-Boltzmann constant. In addition, for a $d$-dimensional perfect black-body emitter, there exists a relation \cite{Bekenstein}
\begin{eqnarray}
 \dot{S}=\frac{d+1}{d}\times \frac{P}{T}.\label{stt}
\end{eqnarray}
For a $d$-dimensional Schwarzschild black hole, its temperature is given by \cite{Tangherlini}
\begin{eqnarray}
 T=\frac{(d-3)\hbar}{4\pi r_{h}}.
\end{eqnarray}
The parameter $r_{h}$ is the radius for the black hole event horizon. While the effective area in Eq. (\ref{pat}) has different interpretations. Mirza, Oboudiat, and Zare evaluated it as the area of black hole horizon, i.e., $\mathcal{A}=A_{h}$. Hod adopted another expression
\begin{eqnarray}
\mathcal{A}=\frac{\Gamma(\frac{d}{2})}{\sqrt{\pi}(d-1)\Gamma(\frac{d-1}{2})}
            \left(\frac{d}{2}\right)^{\frac{d-1}{d-2}}
            \left(\frac{d}{d-2}\right)^{\frac{d-1}{2}} A_{h}.
\end{eqnarray}
Nevertheless,
\begin{eqnarray}
 \mathcal{A}\propto A_{h}.\label{ahh}
\end{eqnarray}
Making use of Eqs. (\ref{pat})-(\ref{ahh}), one will get
\begin{eqnarray}
 \dot{S}\propto P^{\frac{1}{2}},
\end{eqnarray}
which implies that a $d$-dimensional Schwarzschild black hole is also a one-dimensional entropy emitter.

\section{Our argument}

\begin{figure}
\center{\includegraphics[width=6.5cm]{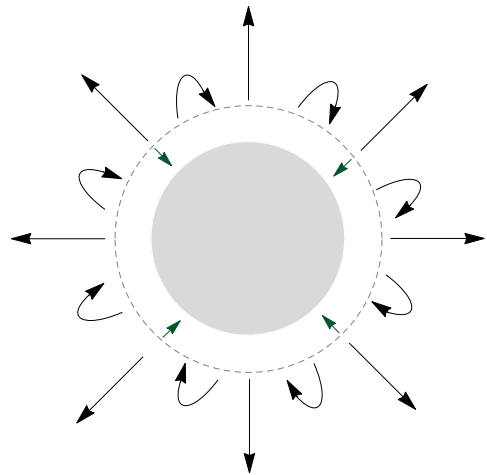}}
\caption{Sketch picture for the black hole radiation. The outward arrows denotes the quanta emitted from the black hole. The twisted arrows represent the quanta reflected by the gravitational potential. The short inward arrows mean that during the quanta emission, the size of the black hole decreases from the region bound by the dashed circle to the shadow region.}\label{rhat1}
\end{figure}

During the treatment, one important key is the relation (\ref{stt}) between the entropy emitted rate and the energy power. However, when we replace the black body with a black hole, we should consider the gravitational effect of the black hole when it emits particles. It will lead to the deviation of black hole from a black body. Therefore, the grey-body factor should be included in. Moreover, the black hole energy will be taken away by the emitted quanta, which will lead to the continuous decrease of the black hole size. We show the sketch picture in Fig. \ref{rhat1}.

Considering these effect, the energy radiation power and entropy emission rate of spin $s$ particles for a Schwarzschild black hole are
\begin{eqnarray}
 P&=&\sum_{j}N_{j}^{(s)}\int_{0}^{\infty} \omega h_{j}^{(s)}(\omega)d\omega,\label{pp}\\
 \dot{S}&=&\sum_{j}N_{j}^{(s)}\int_{0}^{\infty} g_{j}^{(s)}(\omega)d\omega,\label{dsds}
\end{eqnarray}
where $j$ is the angular harmonic index. For scalar, vector, and tensor modes, we respectively have \cite{Rubin}
\begin{eqnarray}
 N_{j}^{(s)}&=&\frac{(2j+d-3)(j+d-4)}{j!(d-3)!},\nonumber\\
 N_{j}^{(s)}&=&\frac{j(j+d-3)(2j+d-3)(j+d-5)!}{(j+1)!(d-4)!},\nonumber\\
 N_{j}^{(s)}&=&\frac{(d-4)(d-1)(j+d-2)(j-1)}{2(j+1)!(d-3)!}\nonumber\\
  &&\times (2j+d-3)(j+d-5)!.\nonumber
\end{eqnarray}
The integrands are given by \cite{Wald,Hawkingd}
\begin{eqnarray}
 h_{j}^{(s)}(\omega)&=&\frac{1}{2\pi}\frac{|A_{j}^{(s)}|^{2}}{e^{\frac{\omega}{T}}-(-1)^{2s}},\nonumber\\
 g_{j}^{(s)}(\omega)&=&
   \frac{1}{2\pi}\bigg[\frac{|A_{j}^{(s)}|^{2}}{e^{\frac{\omega}{T}}-(-1)^{2s}}\nonumber\\
      &&\times\ln\bigg(\frac{e^{\frac{\omega}{T}}-(-1)^{2s}}{|A_{j}^{(s)}|^{2}}+(-1)^{2s}\bigg)\nonumber\\
       &&+(-1)^{2s}\ln\bigg(1+\frac{(-1)^{2s}\times|A_{j}^{(s)}|^{2}}{e^{\frac{\omega}{T}}-(-1)^{2s}}\bigg)
       \bigg].\nonumber
\end{eqnarray}
Here $|A_{j}^{(s)}|^{2}$ is the dimensionless greybody factor, which is a function of $\omega$ and the black hole horizon radius, i.e. $|A_{j}^{(s)}|^{2}=|A_{j}^{(s)}|^{2}(\omega, r_{h})$. Or, more explicitly, we can express the greybody factor as $|A_{j}^{(s)}|^{2}=|A_{j}^{(s)}|^{2}(\omega r_{h})$ with $\omega r_{h}$ being a dimensionless parameter. For the $d$-dimensional Schwarzschild black hole, we have $T\propto \frac{1}{r_{h}}$. Then we can define a new dimensionless variable
\begin{eqnarray}
 x=\frac{\omega}{T}.
\end{eqnarray}
Then the integrals (\ref{pp}) and (\ref{dsds}) will be
\begin{eqnarray}
 P&=&T^{2}\times\sum_{j}N_{j}^{(s)}\int_{0}^{\infty} x h_{j}^{(s)}(x)dx,\\
 \dot{S}&=&T\times\sum_{j}N_{j}^{(s)}\int_{0}^{\infty} g_{j}^{(s)}(x)dx,
\end{eqnarray}
The integral parts in the above equations give pure numbers. So, we have$P\propto T^{2}$ and $\dot{S}\propto T$. Combining these two relations yields
\begin{eqnarray}
 \dot{S}=C(d,s)\times P^{\frac{1}{2}}.\label{dd}
\end{eqnarray}
Here the number $C(d,s)$ is only dependent of the dimension number $d$ of the spacetime and the kind of the emitted particles, while independent of the black hole parameter. It can be obtained by performing the integrals,
\begin{eqnarray}
 C(d,s)=\sum_{j}\sqrt{N_{j}^{(s)}}\frac{\int_{0}^{\infty} g_{j}^{(s)}(x)dx}
 {\sqrt{\int_{0}^{\infty} x h_{j}^{(s)}(x)dx}}.
\end{eqnarray}
During the radiation, the black hole horizon shrinks, which leads to the rise of the black hole temperature. However, the relation (\ref{dd}) always holds during the process. Thus, a $d$-dimensional Schwarzschild black hole is a one-dimensional entropy emitter, even the gravitational effective is included in. It is also clear that such property is independent of the nature of the emitted quanta.

\section{General argument}
\label{argument}

It was reported in Refs. \cite{Mirza,Hod} that, some black holes, such as charged RN black holes, Lovelock black holes, are not one-dimensional entropy emitters. Here we would like to examine the entropy emitter for a general case using the method of dimensional analysis. We adopt the units $k_{B}=G=c=\hbar=1$. For a black hole, the radius of its event horizon is a characteristic length, so we can express all the parameters with the length dimension $[L]$. Taking an example, the black hole temperature $T=T(r_{h}, \alpha_{i})$ with $\alpha_{i}$ being the black hole characteristic parameters (except the mass), such as the charge and angular momentum. Then we can construct a dimensionless temperature $\tilde{T}=r_{h}T(\tilde{\alpha_{i}})=\tilde{T}(\tilde{\alpha_{i}})$, where $\tilde{\alpha_{i}}$ has been non-dimensionalized with $r_{h}$.

For a black hole, the temperature, energy radiation power, and entropy emission rate have the following dimensions
\begin{eqnarray}
 [T] &=& [L]^{-1},\\ ~
 [P] &=& [T^2]=[L]^{-2} ,\\
   \dot{[S]} &=& [T]=[L]^{-1}.
\end{eqnarray}
Then, according to the dimensions, we finally get
\begin{eqnarray}
 \dot{S}= C(d,s;\tilde{\alpha}_{i})\times P^{\frac{1}{2}}.\label{last}
\end{eqnarray}
At a first glance, one will obtain the result that it is the same functional relation as the one-dimensional entropy emitter. However, we should note the expression of the parameters $\tilde{\alpha}_{i}$:
\begin{eqnarray}
 \tilde{\alpha}_{i}=\frac{\alpha_{i}}{r_{h}^{\delta_{i}}},
\end{eqnarray}
where $\delta_{i}$ is the dimensional number of the black hole parameter $\alpha_{i}$. We need clearly recall that, during the radiation of the quanta, the characteristic parameters $\alpha_{i}$ keeps constant, while the horizon shrinks. So the dimensionless parameters $\tilde{\alpha}_{i}$ varies in the emitted process. And the coefficient $C(d,s;\tilde{\alpha}_{i})$ can not be a constant. Therefore, for a black hole with nonvanishing $\alpha_{i}$, it will not a one-dimensional black hole entropy emitter.

On the other hand, a Schwarzschild black hole behaves as a limit for other black holes. Thus in the limit $\alpha_{i}\rightarrow0$, $C(d,s;\tilde{\alpha}_{i})\rightarrow C(d,s)$, then the property of one-dimensional entropy emitter will be recovered.

In fact, if the emitted quanta can change the black hole parameters (like the charge and angular momentum) a specific manner such that $\tilde{\alpha}_{i}$ keep as fixed constants, then the black hole will be exactly a one-dimensional entropy emitter.

\section{Summary}

At last, we would like to give a brief summary of this work. We considered the gravitational effect of black hole on the entropy emission properties.

For a $d$-dimensional Schwarzschild black hole, our result confirms that of Refs. \cite{Mirza,Hod0}. Moreover, we also showed that the one-dimensional entropy emission property of a $d$-dimensional Schwarzschild black hole is independent of the spin $s$ of the emitted quanta. This result holds both for boson and fermion emitted quanta.

When other parameters are included in, for example, the black hole charge and angular momentum, the black hole entropy emission will deviate from the one-dimensional property. This result also agrees with Refs. \cite{Mirza,Hod}.

In summary, a Schwarzschild black hole is a one-dimensional entropy emitter for any spacetime dimension $d$ and for any emitted quanta with spin $s$. While other black holes will not hold such intriguing property in general.

\section*{Acknowledgements}
This work was supported by the National Natural Science Foundation of China (Grants No. 11675064, No. 11522541, No. 11375075, and  No 11205074), and the Fundamental Research Funds for the Central Universities (No. lzujbky-2016-121 and lzujbky-2016-k04).

\end{document}